\documentclass[11pt, onecolumn, draftcls, singlespace ]{IEEEtran}
\usepackage{cite}
\usepackage{algorithm2e}
\usepackage{amssymb}
\usepackage{enumerate}
\usepackage{graphicx}
\usepackage{amssymb}
\usepackage{amsbsy}
\usepackage{float}
\usepackage{balance}
\usepackage{url}
\usepackage{bm}
\usepackage{array}
\usepackage{varioref}
\usepackage[normalem]{ulem}
\graphicspath{{./figs/}}
\usepackage{color}
\usepackage{empheq}
\usepackage{flushend}
\usepackage{amsmath}

\begin{document}

\title{ Multicast Multigroup Beamforming for Per-antenna Power Constrained Large-scale Arrays }
\author{\IEEEauthorblockN{ Dimitrios Christopoulos\IEEEauthorrefmark{1},   Symeon Chatzinotas\IEEEauthorrefmark{1}
 and Bj\"{o}rn Ottersten\IEEEauthorrefmark{1}\\
 \thanks{This work was   partially supported by the National Research Fund, Luxembourg under the project  ``$\mathrm{ CO^{2}SAT}$: Cooperative \& Cognitive Architectures for Satellite Networks'.}
 }

\IEEEauthorblockA{\IEEEauthorrefmark{1}SnT - securityandtrust.lu,  University of Luxembourg
\\email: \textbraceleft dimitrios.christopoulos, symeon.chatzinotas, bjorn.ottersten\textbraceright@uni.lu}

%
}
\maketitle


\begin{abstract}
Large in the number of transmit elements, multi-antenna arrays  with per-element limitations are in the focus of the present work.  
In this context,    physical layer multigroup multicasting under per-antenna power constrains, is investigated herein.   To address this complex  optimization problem low-complexity alternatives to semi-definite relaxation are proposed.   The  goal is to optimize the per-antenna power constrained  transmitter  in a  maximum fairness sense, which is  formulated as a non-convex quadratically constrained quadratic problem. Therefore,  the recently developed tool of feasible point pursuit and successive convex approximation is extended   to account for practical per-antenna power constraints.  Interestingly, the novel iterative method exhibits not only superior performance in terms of approaching the relaxed upper bound  but also a significant complexity reduction, as the dimensions of the optimization variables increase. Consequently, 
 multicast multigroup beamforming for  large-scale array transmitters with per-antenna dedicated amplifiers is rendered computationally efficient and accurate.     A preliminary performance evaluation in large-scale systems for which the semi-definite relaxation constantly yields non rank-1 solutions is presented. \end{abstract}
 \section{{Introduction \& Related Work}}
 Highly demanding applications (e.g. video broadcasting) stretch the throughput limits  of multiuser broadband systems. To provide for such requirements, the adaptation of   the physical layer design of next generation multi-antenna wireless communication systems  to the needs of the higher network  layers is imminent. 
In this direction, physical layer ($\mathrm{PHY}$) multicasting
has the potential to efficiently address the nature of future traffic demand and  has become part of the new generation of communication standards. In-line with the recent trends for spectrally efficient massive multiple input multiple output ($\mathrm{MIMO}$) wireless systems \cite{Rusek2013}, the topic of multicasting over large-scale antenna arrays arises. A brief review of the state-of-the art in multicasting follows. 
\subsection{PHY Multicasting}
The NP-hard multicast problem was defined and accurately approximated by semi-definite relaxation ($\mathrm{SDR}$) and Gaussian randomization in \cite{Sidiropoulos2006}.
Extending the multicast concept, a unified framework for physical layer multicasting to multiple co-channel groups, where independent sets of common data are transmitted to  groups of users by the multiple antennas, was given in \cite{Karipidis2005CAMSAP,Karipidis2008}.  In parallel to \cite{Karipidis2005CAMSAP}, the work   of \cite{Gao2005}  involved  dirty paper coding methods that are bound to increase the complexity of the system. Next, a convex approximation method for the $\max \min$ \textit{fair } optimization was proposed in \cite{Pesavento2012b}, exhibiting increased performance as the number of users per group grows, but for relatively low numbers of transmit antennas. In the same context, a similar iterative convex approximation method, this time for the total power minimization under quality-of-service ($\mathrm{QoS}$) constraints formulation,  was considered in \cite{Bornhorst2011_SPAWC}. In this case, the conservative convex approximation of \cite{Chen2009_ICASSP} was employed and a channel phase based, user scheduling method was performed as a second step towards increasing the tightness of the approximation.   Finally, in \cite{Silva2009},
the multicast multigroup problem, was solved based on approximations and uplink-downlink duality.
{

The hitherto reviewed literature on multigroup multicast beamforming has only considered sum-power constraints  ($\mathrm{SPC}$s) at the transmitter side. 
Amid this extensive literature, the optimal multigroup multicast precoders when a maximum limit is imposed on the  transmitted power of each antenna, have only recently been derived in \cite{Christopoulos2014_ICC,Christopoulos2014_TSP}.   {Therein,  a consolidated solution for the weighted max--min fair multigroup multicast beamforming problem under per-antenna constraints ($\mathrm{PAC}$s)   is  presented.  This framework is based on $\mathrm{SDR}$ and Gaussian randomization  to solve the $\mathrm{QoS}$ problem and bisection to derive an accurate approximation of the non-convex $\max \min$ \textit{fair }formulation. However, as detailed  in \cite{Christopoulos2014_TSP}, the $\mathrm{PAC}$s are bound to increase the complexity of the optimization problem and reduce the accuracy of the approximation, especially as the number of transmit antennas is increasing. These observations necessitate the investigation of lower complexity, accurate approximations that can be applied on large-scale antenna arrays, constrained by practical, per-antenna power limitations. 
\subsection{Successive Convex Approximation}
 Inspired by the recent development of the feasible point pursuit ($\mathrm{FPP}$) successive convex approximation ($\mathrm{SCA}$) of non-convex quadratically constrained quadratic problems ($\mathrm{QCQP}$s), as developed in\cite{Sidiropoulos2015_SPL}, the present work aims at improving the $\max \min$ \textit{fair }  solutions of \cite{Christopoulos2014_TSP}. The $\mathrm{FPP-SCA}$  tool  has been preferred over other existing approximations (for instance \cite{Tran2015_SPL}) due to its guaranteed feasibility regardless of the initial state of the iterative optimization \cite{Sidiropoulos2015_SPL}.

The rest of the paper is structured as follows.   The generic per-antenna power constrained multicast multigroup  system model is presented in Sec. \ref{sec: System Model} while the $\max \min$ problem is formulated and  solved in Sec. \ref{sec: problem}. In Sec. \ref{sec: performance}, the performance of the design is evaluated for a specific system setup. Finally,  Sec. \ref{sec: conclusions} concludes the paper.

{\textit{Notation}: In the remainder of this paper, bold face lower case and upper case characters denote column vectors  and matrices, respectively. The operators \(\left(\cdot\right)^\dag\), $|\cdot|$ and $\otimes$   correspond to  the conjugate transpose,  the absolute value and the Kronecker product respectively,  while $[\cdot]_{ij}  $  denotes the $i, j$-th element of a matrix. An identity matrix of $N\times N $ dimensions is denoted as $\mathbf I_N$ and its $k$-th column as $\mathbf e_k$.  Calligraphic indexed characters denote sets}.
{$\mathbb R_{M}^+$ denotes the set of real positive $M$-dimensional vectors.} 


\section{System Model }\label{sec: System Model}
 Assuming a single transmitter, let   $N_t$ denote the number of transmitting elements  and  $N_{u}$ the  total number of users served.  The input-output analytical expression  will read as $y_{i}= \mathbf h^{\dag}_{i}\mathbf x+n_{i},$
where \(\mathbf h^{\dag}_{i}\) is a \(1 \times N_{t}\) vector composed of the channel coefficients (i.e. channel gains and phases) between the \(i\)-th user and the  \(N_{t}\) antennas of the transmitter, \(\mathbf x\) is the \(N_{t} \times 1\)  vector of the transmitted symbols and  \(n_{i}\) is the independent complex circular symmetric (c.c.s.) independent identically distributed (i.i.d) zero mean  Additive White Gaussian Noise ($\mathrm{AWGN}$)  measured at the \(i\)-th user's receive antenna.
Focusing  on a multigroup multicasting scenario,  let there be a total of $1\leq G \leq N_{u}$ multicast groups with  $\mathcal{I} = \{\mathcal{G}_1, \mathcal{G}_2, \dots  \mathcal{G}_G\}$ the collection of   index sets and $\mathcal{G}_k$ the set of users that belong to the $k$-th multicast group, $k \in \{1\dots G \}$. Each user belongs to only one group, thus $\mathcal{G}_i\cap\mathcal{G}_j=$\O ,$  \forall i,j \in \{1\cdots G\}$. Let $\mathbf w_k \in \mathbb{C}^{N_t \times 1}$ denote the precoding weight vector applied to the transmit antennas to beamform towards the $k$-th group.
The assumption of independent data transmitted to different groups renders the symbol streams $\{s_k\}_{k=1}^G$ mutually uncorrelated and the total power radiated from the antenna array is
$P_{tot} = \sum_{k=1}^ G \mathbf w_k{^\dag} \mathbf w_k$. The power radiated by each
antenna element is  a  linear combination of all precoders $P_n = \left[\sum_{k=1}^G \mathbf w_k \mathbf w_k^\dag \right]_{nn}$,
where $n \in \{1\dots  N_t\}$ is the antenna index.

\section{Multicast Multigroup under $\mathrm{PAC}$s}\label{sec: problem}
\subsection{SDR Based Solution}
\subsubsection{ Max-Min Fair Formulation}

\begin{empheq}[box=\fbox]{align}
\mathcal{F:}\    \max_{\  t, \ \{\mathbf w_k \}_{k=1}^{G}}  &t& \notag\\
\mbox{subject to } & \frac{1}{\gamma_i}\frac{|\mathbf w_k^\dag \mathbf h_i|^2}{\sum_{{l\neq k }}|\mathbf w_l^\dag\mathbf h_i|^2+\sigma_i^2 }\geq t, &\label{const: F SINR}\\
&\forall i \in\mathcal{G}_k, k, l\in\{1\dots G\},\notag\\
 \text{and to }\ \ \ \ & \left[\sum_{k=1}^G  \mathbf w_k\mathbf w_k^\dag  \right]_{nn}  \leq P_n, \label{eq: max-min fair power const1 }\\
 &\forall n\in \{1\dots N_{t}\},\notag
 \end{empheq}
 where $\mathbf w_k\in \mathbb{C}^{N_t}$ and $t \in \mathbb{R}^{+}$. {The notation $\sum_{l \neq k}$ states that aggregate interference from all co-channel groups is calculated.}   Problem $ \mathcal{F}$ receives as inputs the  $\mathrm{PAC}$s vector $\mathbf p = [P_1, P_2\dots P_{N_t}]$ and the target $\mathrm{SINR}$s vector $\mathbf g = [\gamma_1,\gamma_2, \dots \gamma_{N_u}]$. {Its goal is to maximize the slack variable $t$ while keeping all $\mathrm{SINR}$s above this value. Thus, it constitutes a max-min problem that guarantees fairness amongst users.}    The main complication of problem $ \mathcal{F}$ lies in constraint \eqref{const: F SINR}, where a multiplication of the two optimization variables takes place. To reduce this formulation into the more tractable   $\mathrm{QCQP}$ form, the following considerations are emanated.

\subsubsection{Per-antenna Power  Minimization}
{ A relation between the fairness  and the power minimization problems for the multicast multigroup case under $\mathrm{SPC}$s was firstly established in \cite{Karipidis2008}. As a result, by bisecting  the solution of the  $\mathrm{QoS}$ optimization, a solution to the weighted fairness problem can be derived. Nevertheless, fundamental differences between the $\mathrm{SPC}$ formulation and the $\mathrm{PAC}$ problem $\mathcal{F}$, complicate the solution. In more detail, the PACs --i.e \eqref{eq: max-min fair power const1 }-- are not necessarily met with equality. A more detailed discussion on this can be found in \cite{Christopoulos2014_TSP}. Therefore,  a  per-antenna power  minimization problem has been proposed in \cite{Christopoulos2014_TSP},  as}
\begin{empheq}[box=\fbox]{align}
\mathcal{Q:} \min_{\ r, \ \{\mathbf w_k \}_{k=1}^{G}}  &r& \notag\\
\mbox{subject to } & \frac{|\mathbf w_k^\dag \mathbf h_i|^2}{\sum_{l\neq k } | \mathbf w_l^\dag\mathbf h_i|^2 + \sigma^2_i}\geq \gamma_i, \label{const: Q SINR}\\
&\forall i \in\mathcal{G}_k, k,l\in\{1\dots G\},\notag\\
\text{and to} \ \ \ \ \ & \frac{1}{P_n} \left[\sum_{k=1}^G  \mathbf w_k\mathbf w_k^\dag \right]_{nn} \leq  r,\label{eq: PAC Q}\\
& \forall n\in \{1\dots N_{t}\}, \notag
 \end{empheq}
 with $r\in\mathbb{R^+}$. Problem $\mathcal{Q }$ receives as input  $\mathrm{SINR}$ constraints for  all users, defined   before as $\mathbf g $,  as well as the per antenna power constraint vector $\mathbf p$ of \eqref{eq: max-min fair power const1 }. {The introduction of the slack-variable  $r$,   constraints the power consumption of each and every antenna.   Subsequently, at the optimum $r^*$,   the maximum  power consumption out of all antennas is minimized} and this solution is denoted as $r^*=\mathcal{Q}(\mathbf g, \mathbf p)$. 

 \textit{Claim 1}: Problems $\mathcal{F}\ $ and  $\mathcal{Q}\ $ are related as follows
\begin{align}
1 = \mathcal{Q}\left(\mathcal{F}\left( \mathbf g, \mathbf p\right)\cdot\mathbf g, \mathbf p \right)\label{eq: equivalence 1}\\
t = \mathcal{F}\left(\mathbf g, \mathcal{Q}\left( t\cdot \mathbf g,\mathbf p\right)\cdot\mathbf p \right)\label{eq: equivalence 2}
 \end{align}
(for {proof } cf. \cite{Christopoulos2014_TSP}) $\blacksquare$
\subsubsection{Bisection}\label{sec: bisec}

{The establishment of claim 1  allows for the application of the bisection method, as developed in \cite{Sidiropoulos2006,Karipidis2008}.} The solution of $
r^* = \mathcal{Q}_r \left ( \frac{L+U}{2}\mathbf g, \mathbf p\right)
$ is obtained by bisecting the interval $[L, U]$  as defined by the minimum and maximum $\mathrm{SINR}$ values. Since $t=(L+U)/2$ represents the $\mathrm{SINR}$
, it will always be positive or zero. Thus, $L = 0.$ Also, if the system was interference  free while all the users had the channel of the best user, then the maximum worst $\mathrm{SINR}$ would be attained, thus $U = \max_i\{P_{tot}\mathbf Q_i/\sigma_i \}. $ If $r^*<1$,   then the lower bound of the interval is updated with this value. Otherwise the value is assigned to the upper bound of the interval. Bisection is iteratively performed until an the interval size is reduced to a pre-specified value $\epsilon$ (herein, $\epsilon = 10^{-3}$).
This value needs to be dependent on the magnitude of $L \text{ and } U$ so that the accuracy of the solution is maintained regardless of the region of operation.
 After a finite number of iterations, the optimal value of $\mathcal{F}$ is given as the resulting value for which $L \text{ and } U$ become almost identical, providing an accurate solution for $\mathcal{F}$.
\subsubsection{Relaxation and Gaussian Randomization} The bisection method, as previously discussed, overcomes the non-convexity due to the multiplication of two variables, namely $t$ and $\mathbf w$ in constraint \eqref{const: F SINR}. However,  problem $\mathcal Q$ still remains non-convex.  Based on the observation that
$|\mathbf w_k^\dag \mathbf h_i|^2 = \mathbf w_k^\dag \mathbf h_i \mathbf h_i^\dag \mathbf w_k = {\mathrm{Tr}}(\mathbf w_k^\dag \mathbf h_i \mathbf h_i^\dag \mathbf w_k) = {\mathrm{Tr}}(\mathbf w_k \mathbf w_k^\dag \mathbf h_i \mathbf h_i^\dag )$ and with the change of variables $\mathbf X_i = \mathbf w_i \mathbf w_i^\dag $, one can easily identify that the non-convexity of 
 $\mathcal{Q}$ lies in the necessity to constrain variable $\mathbf X $ to have a unit rank. By dropping this constraint, the non-convex $\mathcal Q$ can be relaxed to
 $\mathcal{Q}_r$, which reads as
 \begin{empheq}[box=\fbox]{align}
\mathcal{Q}_r:\min_{r,\ \{\mathbf X_k \}_{k=1}^{G}}  &r& \notag\\
\mbox{subject to } & \frac{\mathrm{Tr}\left(\mathbf h_{i}\mathbf h^\dag_{{i}} \mathbf X_k\right)}{\sum_{l\neq k } \mathrm{Tr}\left(\mathbf h_{i}\mathbf h^\dag_{{i}} \mathbf X_{{l}}\right) +\sigma_i^2}\geq \gamma_i, \label{const: Q_r SINR}\\
&\forall i \in\mathcal{G}_k, k,l\in\{1\dots G\},\notag\\
 \text{and to} \ \ \ \ \ &\frac{1}{P_n} \left[\sum_{k=1}^G \mathbf X_{k}\right]_{nn} \leq  r   \label{const: Q_r Power}\\
 &\forall n\in \{1\dots N_{t}\}, \notag\\
 \text{and to}  \ \ \ \ \ & \ \mathbf X_{k}\succeq 0,\label{const: Q_r SDF} \ {\forall k\in \{1\dots G\}},
 \end{empheq}
 Following this relaxation, the derivation of the optimal value  $\mathbf w^{*}$ requires a rank-1 approximation over  $\mathbf X ^*$. The approximation with the highest accuracy is proven to be the Gaussian approximation \cite{Luo2010}. In summary, this procedure involves the generation of  precoding vectors drawn from a Gaussian  distribution  with statistics defined by the relaxed solution. After generating a a number of instances and re-scaling them, the solution with the closest performance to the relaxed upper bound, as given by the optimal point of $\mathcal{Q}_r$ is chosen.     More details on the $\mathrm{SDR}$ based solution  under PACs, can be found in \cite{Christopoulos2014_TSP}.
 \subsection{Successive Convex Approximation}\label{sec: SDR}
  Problem  $\mathcal{Q}$ belongs in the general class of non-convex $\mathrm{QCQP}$s for which the $\mathrm{SDR}$ technique is proven to be a powerful and computationally efficient approximation technique \cite{Luo2010}. However, the $\mathrm{FPP-SCA}$, a recently proposed alternative to $\mathrm{SDR}$, is herein considered \cite{Sidiropoulos2015_SPL}.
By defining $\mathbf w_{tot} = [\mathbf w_1^\dag, \mathbf w_2^\dag \dots \mathbf w_G^\dag]^\dag$, the $i$-th $\mathrm{SINR}$ constraint reads as
\begin{align}
\mathbf w_{tot}^\dag \mathbf A_{i} \mathbf w_{tot}\leq -\gamma_i\sigma_i^2, \label{const: Q SCA SINR}
\end{align}
where  $\mathbf A_i = \mathbf A_i^{(+)}+ \mathbf A_i^{(-)}$ with  $  \mathbf A_i^{(+)}= \gamma_i\left(\mathbf I _G - \text{diag}\{\mathbf e_{k}\}\right)\otimes \mathbf h_i\mathbf h_i ^\dag  $ and $  \mathbf A_i^{(-)}=  - \text{diag}\{\mathbf e_{k}\}\otimes \mathbf h_i\mathbf h_i ^\dag  $ , $\forall i \in\mathcal{G}_k$. Assuming a random point $\mathbf z$, then by the definition of a semi-definite matrix $\mathbf A_i^{(-)}$ we have   $\left(\mathbf w_{tot} - \mathbf z \right)^\dag\mathbf A_i^{(-)}\left(\mathbf w_{tot} - \mathbf z \right)\leq 0 $. By expanding this,  a linear restriction of $\mathbf w_{tot}$ around $\mathbf z$ reads as\begin{align}
\mathbf w_{tot} ^\dag \mathbf A_i^{(-)} \mathbf w_{tot} \leq 2\mathrm{Re}\left\{\mathbf z ^\dag\mathbf A_i^{(-)}\mathbf w_{tot}\right\}  - \mathbf z ^\dag\mathbf A_i^{(-)}\mathbf z. 
\end{align}
  Consequently, the  $\mathrm{SINR}$ constraint \eqref{const: Q SCA SINR} can be replaced by 
\begin{align}
\mathbf w_{tot} ^\dag \mathbf A_i^{(+)} \mathbf w_{tot} + 2\mathrm{Re}\left\{\mathbf z ^\dag\mathbf A_i^{(-)}\mathbf w_{tot}\right\}  - \mathbf z ^\dag\mathbf A_i^{(-)}\mathbf z \notag   \leq -\gamma_i\sigma_i^2,
\end{align}
in which the unknown variables are quadratic over a semi-definite matrix.
By adding slack penalties $\mathbf s\in \mathbb{R}_{(N_u+1)}^{+}$, the the original $\mathrm{QCQP}$ problem $\mathcal{Q}$ can be approximated by
\begin{empheq}[box=\fbox]{align}
\mathcal{Q}_{SCA}:&\min_{r,  \mathbf w_{tot}, \mathbf s }  r + \lambda ||\mathbf s || \notag\\
\mbox{s.t.}& \ \mathbf w_{tot} ^\dag \mathbf A_i^{(+)} \mathbf w_{tot} + 2\mathrm{Re}\left\{\mathbf z^{(j)\dag} \mathbf A_i^{(-)}\mathbf w_{tot}\right\} \\& - \mathbf z^{(j)\dag} \mathbf A_i^{(-)}\mathbf z^{(j)} \notag   \leq -\gamma_i\sigma_i^2 + s_i\\
&\forall i \in\mathcal{G}_k, k,l\in\{1\dots G\},\notag\\
 \text{and to}& \ \frac{1}{P_n} \left[ \mathbf w_{tot} \mathbf w_{tot}^\dag\right]_{nn} \leq  r+s_{N_u+1}   \label{const: Q_r Power}\\
 &\forall n\in \{1\dots N_{t}\}, \notag
 \end{empheq}
where $r \in \mathbb{R}^{+}$, $\lambda \in\mathbb{R}$ is a fixed input parameter and $\mathbf z^{(j)}$ is the $j-$th instance of the introduced auxiliary variable. In each instance of the $\mathrm{SCA}$ algorithm, $\mathcal Q_{SCA}$ is solved and the starting point is updated as $\mathbf z^{(j+1)} = \mathbf w_{tot}^{(j)}$. The iterative process is repeated until the guaranteed convergence \cite{Sidiropoulos2015_SPL}. 
\subsection{Complexity \& Convergence discussions}\label{sec: complexity}
 An important discussion involves the complexity of the employed techniques  to approximate a solution of  the highly complex, NP-hard  multigroup multicast problem under $\mathrm{PAC}$s.   Focusing on the  $\mathrm{SDR}$ based  solution of \cite{Christopoulos2014_TSP}, the main complexity burden originates from the relaxed $\mathcal{Q}_r$.       The total worst case complexity of the $\mathrm{SDR}$ based solution of  $\mathcal{F}$, as in detail is calculated in  \cite{Christopoulos2014_TSP}, is summarised in the following.
Initially, a  bisection search is performed over  $\mathcal{Q}_r$ to  obtain the relaxed solution.  This bisection runs for $N_{iter} = \lceil\log_2\left(U_{1}-L_{1}\right)/\epsilon_{1}\rceil$ where $\epsilon_1 $ is the desired accuracy of the search.  Typically   $\epsilon_1 $ needs to be at least three orders of magnitude below the magnitudes of $U_{1}, L_{1}$ for sufficient accuracy. In each iteration of the bisection search, problem $\mathcal{Q}_r$ is solved.  This $\mathrm{SDP}$ has $G$ matrix variables of $N_t\times N_t$ dimensions and $N_{u}+N_t$ linear constraints.    Moreover, in each iteration not more than $\mathcal{O}({G^3 N_t^6 +GN_{t}^{3} +N_{u}GN_t^2})$ arithmetic operations will be performed.      Next,  a fixed number of Gaussian random instances with covariance given by the previous solution are generated. {The complexity of this process is linear with respect to  the number of Gaussian randomizations}. More details on the total complexity of the $\mathrm{SDR}$ based algorithm can be found in \cite{Christopoulos2014_TSP} and are herein omitted for shortness. 

As far as the $\mathrm{FPP-SCA}$ method is concerned, the iterative process typically runs for a few iterations, especially for larger values of $\lambda$. As in detail explained in \cite{Sidiropoulos2015_SPL}, convergence is guaranteed. Therein, $\lambda$ was set to 10 while herein even greater values are chosen, i.e. $\lambda = 25$ since the optimization problems tackled involve a larger number of constraints. Therefore,  In each iteration of the $\mathrm{FPP-SCA}$, bisection search is performed over  $\mathcal{Q}_{SAC}$. The later, is a second order cone program with a worst case complexity of $\mathcal{O}((G N_t+N_u)^{3.5})$. The later fact  justifies the user of the $\mathrm{FPP-SCA}$ in scenarios where the number of transmit antennas exceeds the number of users.

\section{ Performance Evaluation \& Applications} \label{sec: performance}

\subsection{{Uniform Linear Arrays}}
To the end of investigating the sensitivity of the proposed algorithm in a generic environment, a uniform linear array ($\mathrm{ULA}$) transmitter is considered. Assuming far-field, line-of-sight conditions, the user  channels  can be  modeled using Vandermonde matrices. For this important special case, the $\mathrm{SPC}$ multicast multigroup problem was reformulated into a convex optimization problem and solved in \cite{Karipidis2006,Karipidis2007}. These results where motivated by the observation that in sum power constrained  $\mathrm{ULA}$ scenarios, the relaxation consistently yields rank one solutions. Thus, for such cases, the $\mathrm{SDR}$ is essentially optimal \cite{Sidiropoulos2006}. Nevertheless, the $\mathrm{SDR}$ of the $\mathrm{PAC}$ minimization problem in $\mathrm{ULA}$s  is not always tight as shown in \cite{Christopoulos2014_TSP}. 

Let us consider a $\mathrm{ULA}$ serving $4$ users allocated to $2$ distinct groups.  In Fig. \ref{fig: ULA pos}, its radiation pattern for $N_t = 8$ antennas and for co-group angular separation $\theta_a = 35^\circ$  is plotted. A total power budget of $P = -3$~dBW is equally distributed amongst the available antennas. For the Gaussian randomization, $N_\mathrm{rand} = 100$ instances are considered. Clearly, the multigroup multicast beamforming optimizes the lobes  to reduce interferences between the two groups. The beam patterns from both $\mathrm{SDR}$ and $\mathrm{FPP-SCA}$ solutions are included in Fig. \ref{fig: ULA pos}. The superiority in terms of minimum achievable $\mathrm{SINR}$ of the latter solution is apparent.
Hereafter, the performance evaluation  will be based  on the minimum user rate, since in the optimization all users are  equally weighted.
 
Firstly, the performance with respect to the angular separation of co-group users is investigated, as $\theta_a$ is increased for both groups in the fashion indicated in Fig. \ref{fig: ULA pos}. In Fig. \ref{fig: ULA SINR},  when co-group users are collocated, i.e. $\theta_a = 0^\circ$, the highest minimum rate is attained.  As the separation increases, the rate is reduced reaching a local  minimum when interfering users  are placed in the same position, i.e. $\theta_a = 45^\circ$. Then, the lowest value is observed when co-group users are orthogonal, i.e. $\theta_a = 90^\circ$. In Fig. \ref{fig: ULA SINR}, the lack of tightness of the relaxation for the  $\mathrm{SDR}$ based solution is clear as the channel conditions are deteriorating. The only exception is when $\theta_a = 60^\circ$, where the inherent symmetricity of the $\mathrm{ULA}$ transmitter is providing sufficient conditions for a rank-1 solution to be easily obtained. Interestingly, this is the only situation where the $\mathrm{FPP-SCA}$ method provides a suboptimal solution. For all other instances, the superiority of the lower complexity solution is clear.  Consequently, the $\mathrm{FPP-SCA}$ outperforms $\mathrm{SDR}$, over the majority of  the span of the angular separations, for moderately sized $\mathrm{ULA}$s.
In the same setting, the normalized simulation time to compute each precoder is given in Fig. \ref{fig: ULA time 1}. Clearly, when the $\mathrm{SDR}$ does not yield rank-1 solutions, the $\mathrm{FPP-SCA}$ methods can not only provide more accurate solutions but also at a significantly reduced time. Almost 50\% of gains in terms of simulation time are observed at $\theta_a = 80^\circ$.  

 \begin{figure}
 \centering
 \includegraphics[width=0.65\columnwidth]{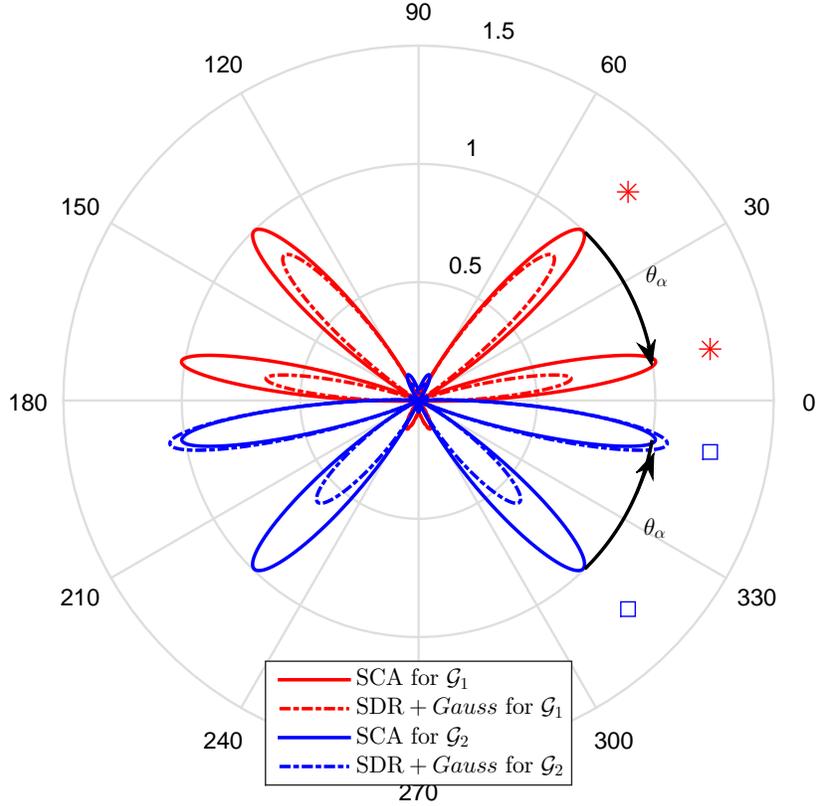}\\
  \caption{ULA beampattern for $\mathrm{PAC}$ and re-scaled $\mathrm{SPC}$ solutions.}
\label{fig: ULA pos}
\end{figure}
 \begin{figure}
 \centering
 \includegraphics[width=0.8\columnwidth]{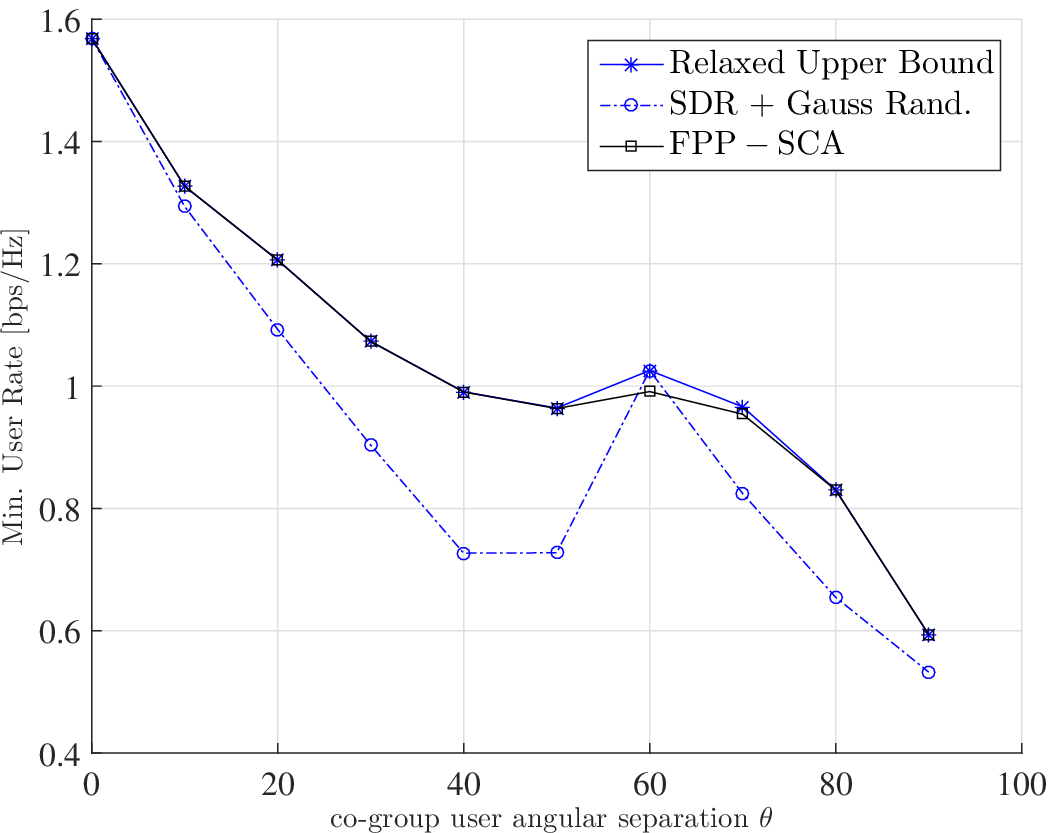}\\
  \caption{ $\mathrm{ULA}$ performance in terms of minimum SINR per group, for increasing co-group user angular separation.}
\label{fig: ULA SINR}
\end{figure}
 \begin{figure}
 \centering
 \includegraphics[width=0.73\columnwidth]{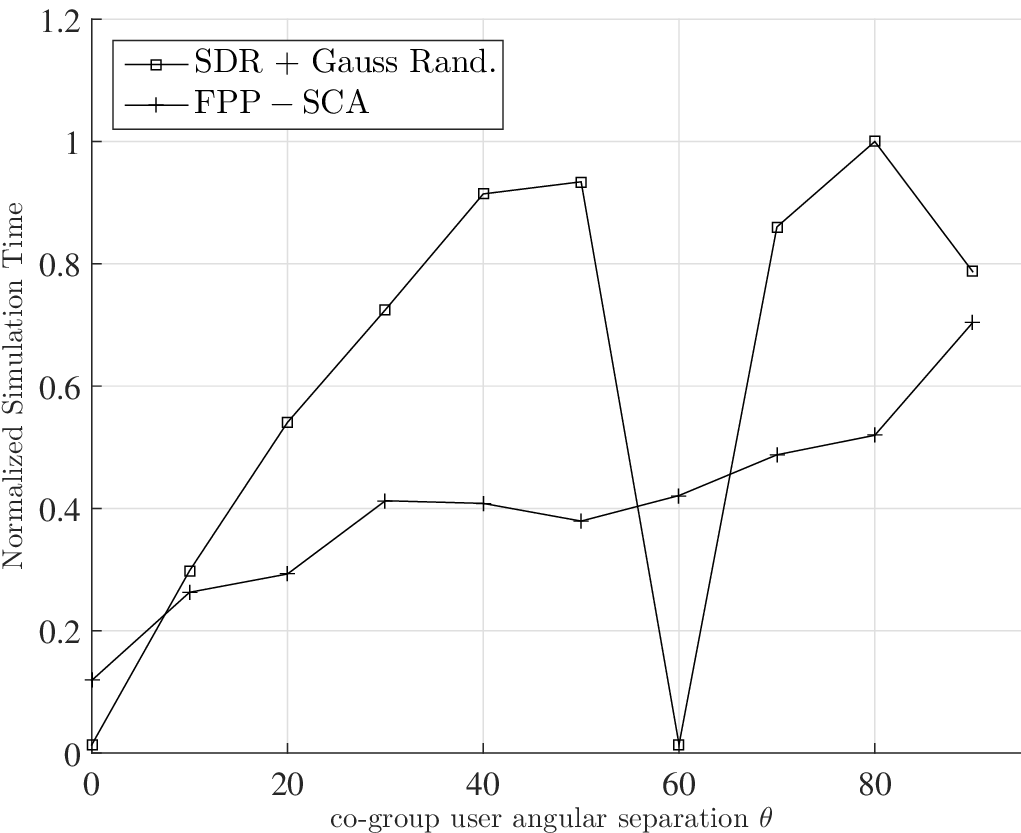}\\
  \caption{ Normalized simulation time for increasing co-group user angular separation.}
\label{fig: ULA time 1}
\end{figure}
Finally, for an angular separation of  $\theta_a = 60^\circ$ where the $\mathrm{FPP-SCA}$ solution performs worse, the minimum rate versus an increasing number of transmit antennas is plotted in Fig.  \ref{fig: ULA SINR 2}, while all other simulation parameters remain unaltered. Therein, the benefits of $\mathrm{FPP-SCA}$ as the number of antennas is increasing are shown. The $\mathrm{SDR}$ solution, fails to provide an accurate solution from 10 antennas onwards. Nevertheless, the $\mathrm{FPP-SCA}$ methods provide a tight approximation to the upper bound irrespective of the number of transmit antennas. Impressively, the almost 20\% of performance gains come also at reduced complexity. As shown in Fig. \ref{fig: ULA time 2}, the simulation time can be reduced by even 80\%, for large-scale antenna arrays. It should be clarified, that the simulation time figures do not follow the complexity dependence given in Sec. \ref{sec: complexity} simply because the considerations mentioned therein involve worst case complexity. Existing solvers employed typically exploit the specific structure of matrices thus reducing the actual execution time.
 \begin{figure}
 \centering
 \includegraphics[width=0.8\columnwidth]{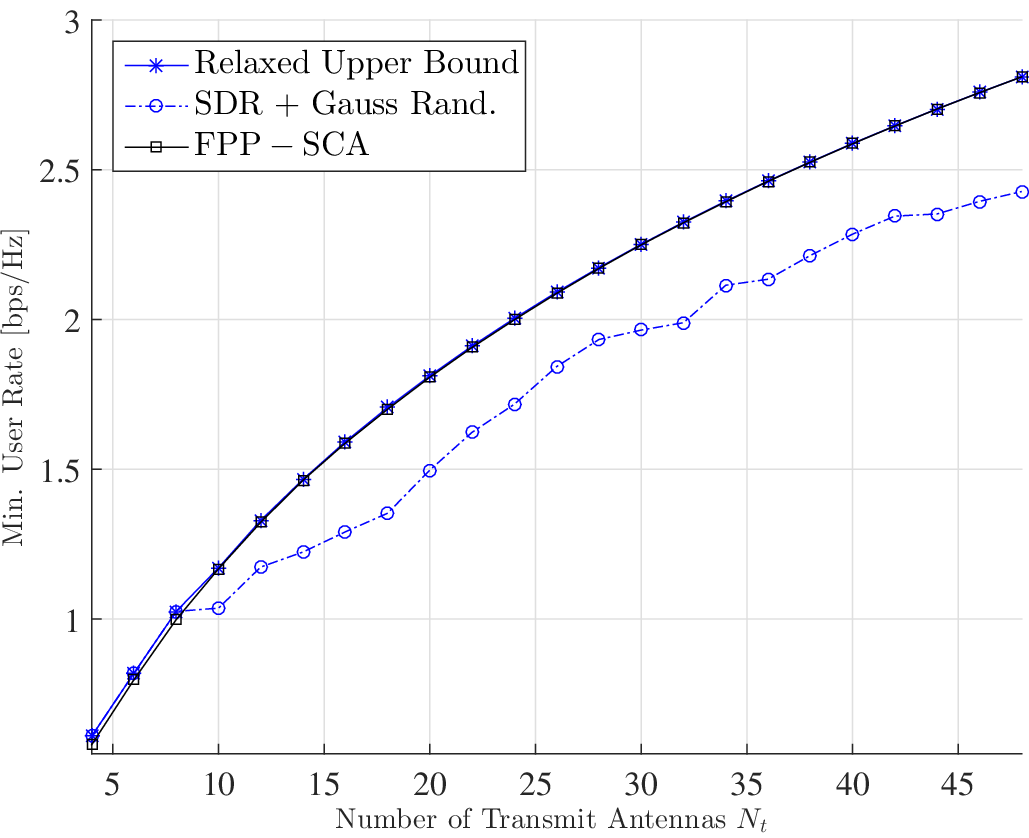}\\
  \caption{ $\mathrm{ULA}$ performance in terms  of minimum user rate versus an increasing number of transmit antennas.}
\label{fig: ULA SINR 2}
\end{figure}
 \begin{figure}
 \centering
 \includegraphics[width=0.8\columnwidth]{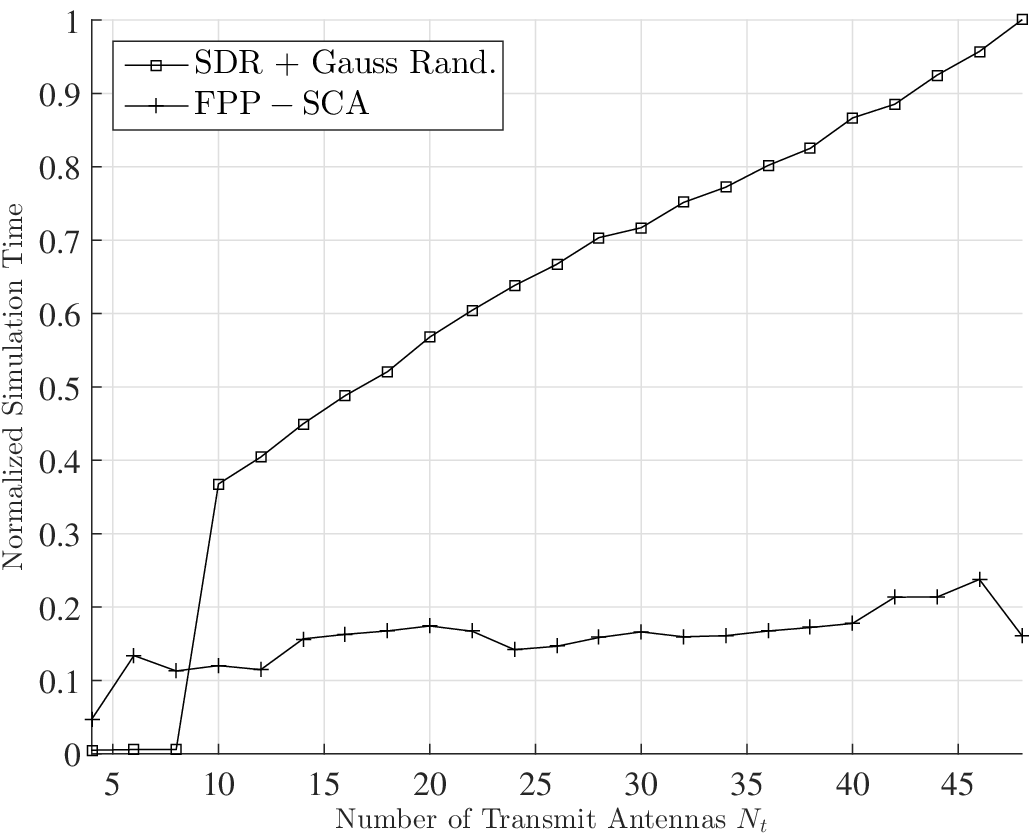}\\
  \caption{ Normalized simulation time versus an increasing number of transmit antennas.}
\label{fig: ULA time 2}
\end{figure}
\section{Conclusions} \label{sec: conclusions}
\textit{}Herein, the  $\max-\min$ \textit{fair} multicast multigroup problem under $\mathrm{PAC}$s  is solved for large-scale antenna arrays. Impressively, the accurate and low complexity  $\mathrm{FPP-SCA}$ methods  outperform existing $\mathrm{SDR}$ based approaches both in terms of complexity as well as accuracy,  as the number of transmit antennas increases. Future extensions of this work involve different optimization criteria such as the sum rate maximization as well as robust formulations.  
%

\textit{Acknowledgements:}
this work was   supported by the research  projects,  $\mathrm{SEMIGOD}$ (National Research Fund, Luxembourg) $\mathrm{SANSA}$ (European Commission  H2020) and $\mathrm{PreDem}$ and $\mathrm{FGBBF}$ (European Space Agency).
\bibliographystyle{IEEEtran}
\bibliography{refs/IEEEabrv,refs/conferences,refs/journals,refs/books,refs/references,refs/csi,refs/thesis}
\end{document}